# Relativistic equilibrium velocity distribution to improve standard solar models

Jian-Miin Liu*

Department of Physics, Nanjing University

Nanjing, The People's Republic of China

*On leave. E-mail address: liu@phys.uri.edu.

ABSTRACT

The current standard solar models can be improved by substituting the relativistic equilibrium velocity distribution for the Maxwellian velocity distribution. The relativistic equilibrium velocity distribution is close-fitting to the Maxwellian velocity distribution in the low-velocity part and substantially different from the Maxwellian velocity distribution in the high-velocity part, and vanishes at the velocity of light. If adopted in standard solar models, it will lower solar neutrino fluxes, change solar neutrino energy spectra but maintain solar sound speeds.

Standard solar models [1-2] deal with our Sun as a typical one of main sequence stars in stellar evolution. They, depending only on several initial inputs, successfully give the Sun's properties such as its total mass, radius, luminosity, temperature and compositions of heavy elements at the Sun age. Some with the diffusion mechanism about helium and heavy elements achieve further success in a very close agreement with the observed helioseimological data [3]. However they all fail to predict solar neutrino signals. This is called the solar neutrino problem [1-4]. The solar neutrino problem contains, besides the difficulty in understanding the observed data of the $^8$B-, $^7$Be-, CNO-, pp- and pep-neutrino fluxes in different experiments from the viewpoint of solar neutrino energy spectra, the discrepancies between the measured solar neutrino fluxes and those predicted by standard solar models. The measured solar neutrino fluxes range from 33%+/-5% to 58%+/-7% of the predicted values [3]. This situation indicates that standard solar models are good but need to be improved.

Solar core produces and radiates its neutrinos primarily through nuclear fusion reactions in the proton-proton cycle and the CNO cycle. To create a nuclear fusion reaction, a proton or a nucleus must penetrate the repulsive Coulomb barrier and collide with another proton or nucleus. As the height of Coulomb barrier is far above solar thermal energy $K_BT_c$: their ratio is typically greater than a thousand [2,4], nuclear fusion reactions can occur only or mostly in those pairs of protons or nuclei having high relative energies. As interacting protons and nuclei in solar interior reach their equilibrium distribution in an infinitesimal period of time compared to the mean lifetime for a nuclear fusion reaction [2,4], an equilibrium velocity distribution is applicable enough to calculating the rates of solar nuclear fusion reactions. In solar interior, therefore, it is the equilibrium velocity distribution of high-energy protons and nuclei that participates in determining solar nuclear fusion reaction rates, and hence, solar neutrino fluxes and solar neutrino energy spectra.

High energy infers high velocity and high velocity is a concept of special relativity. We ought to use the relativistic equilibrium velocity distribution of high-energy protons and nuclei to calculate the rates of solar nuclear fusion reactions. But all standard solar models use the Maxwellian velocity distribution. This is a place, for these well-constructed standard solar models, to be improved because the Maxwellian velocity distribution is obviously inconsistent with special relativity.

In the pre-relativistic mechanics, the velocity space is

$$dY^2 = \delta_{rs} dy^r dy^s, \quad r,s=1,2,3, \tag{1}$$

in the usual velocity-coordinates $\{y^r\}$, $r=1,2,3$, where $y^r$ is the well-defined Newtonian velocity. This velocity space is characterized by (i) unboundedness and (ii) the Galilean addition law.

In the special theory of relativity and the modified special relativity theory [5], the velocity space is

$$dY^2 = H_{rs}(y) dy^r dy^s, \quad r,s=1,2,3, \tag{2a}$$

$$H_{rs}(y) = c^2 \delta^{rs}/(c^2-y^2) + c^2 y^r y^s/(c^2-y^2)^2, \quad \text{real } y^r \text{ and } y<c, \tag{2b}$$

in the usual velocity-coordinates, where $y=(y^r y^r)^{1/2}$, $r=1,2,3$, and c is the velocity of light [6-7]. This velocity space can be represented in the so-called primed velocity-coordinates $\{y'^r\}$, $r=1,2,3$, [6], which are connected with the usual velocity-coordinates by

$$dy'^r = A^r_s(y) dy^s, \quad r,s=1,2,3, \tag{3a}$$

$$A^r_s(y) = \gamma \delta^{rs} + \gamma(\gamma-1) y^r y^s/y^2, \tag{3b}$$

where

$$\gamma = 1/(1-y^2/c^2)^{1/2}. \tag{4}$$

The represented velocity space has the Euclidean structure in the primed velocity-coordinates,

$$dY^2 = \delta_{rs} dy'^r dy'^s, \quad r,s=1,2,3, \tag{5}$$

because

$$H_{rs}(y) = \delta_{pq} A^p_r(y) A^q_s(y), \quad p,q,r,s=1,2,3. \tag{6}$$

Using the calculation techniques in Riemann geometry, at some length, we can find

$$y'^r = \left[\frac{c}{2y} \ln \frac{c+y}{c-y}\right] y^r, \quad r=1,2,3, \tag{7a}$$

$$y' = \frac{c}{2} \ln \frac{c+y}{c-y}, \tag{7b}$$

where $(y'^1, y'^2, y'^3)$ and $(y^1, y^2, y^3)$ denote the same point in the velocity space, and $y'=(y'^r y'^r)^{1/2}$, $r=1,2,3$ [6]. The velocity space embodied in Eqs.(2a-2b) or (5) is characterized by (i) and (ii) in the primed velocity-coordinates, and also by (iii) a finite boundary at c and (iv) the Einstein velocity addition law in the usual velocity-coordinates. We call $y'^r$, $r=1,2,3$, the primed velocity. Its definition from the measurement point of view is given in Ref.[8]. The Galilean addition law of primed velocities links up with the Einstein addition law of corresponding Newtonian velocities [5-6]. Actually, the statement can be easily judged in the one-dimensional case that

$$y'_2 = y'_1 - u' = (c/2)\ln[(c+y_1)/(c-y_1)] - (c/2)\ln[(c+u)/(c-u)]$$

and

$$y'_2 = (c/2)\ln[(c+y_2)/(c-y_2)]$$

imply

$$y_2 = (y_1-u)/[1 - y_1 u/c^2].$$

The Euclidean structure of the velocity space in the primed velocity-coordinates convinces us of the Maxwellian velocity and velocity rate distributions in the primed velocity-coordinates, namely

$$P(y'^1, y'^2, y'^3) dy'^1 dy'^2 dy'^3 = N\left(\frac{m}{2\pi K_B T}\right)^{3/2} \exp\left[-\frac{m}{2K_B T}(y')^2\right] dy'^1 dy'^2 dy'^3 \tag{8}$$

and

$$P(y') dy' = 4\pi N\left(\frac{m}{2\pi K_B T}\right)^{3/2} (y')^2 \exp\left[-\frac{m}{2K_B T}(y')^2\right] dy', \tag{9}$$

where N is the number of particles, m their rest mass, T the temperature, and $K_B$ the Boltzmann constant. Using Eq.(7b) and

$$dy'^1 dy'^2 dy'^3 = \gamma^4 dy^1 dy^2 dy^3$$

which is inferred from Eqs.(3a-3b), we have from Eq.(8),

$$P(y^1, y^2, y^3) dy^1 dy^2 dy^3 = N\frac{(m/2\pi K_B T)^{3/2}}{(1-y^2/c^2)^2} \exp\left[-\frac{mc^2}{8K_B T}\left(\ln\frac{c+y}{c-y}\right)^2\right] dy^1 dy^2 dy^3. \tag{10}$$

Using Eq.(7b) and

$$dy' = \gamma^2 dy$$

which comes from differentiating Eq.(7b), we have from Eq.(9),

$$P(y) dy = \pi c^2 N \frac{(m/2\pi K_B T)^{3/2}}{(1-y^2/c^2)} \left(\ln\frac{c+y}{c-y}\right)^2 \exp\left[-\frac{mc^2}{8K_B T}\left(\ln\frac{c+y}{c-y}\right)^2\right] dy. \tag{11}$$

Eqs.(10) and (11) are the relativistic equilibrium velocity and velocity rate distributions in the usual velocity-coordinates.

$P(y^1, y^2, y^3) dy^1 dy^2 dy^3$ and $P(y) dy$ are so named because they are based on the velocity space in special relativity and the modified special relativity theory. The relativistic equilibrium velocity and velocity rate distributions are close-fitting to the Maxwellian velocity and velocity rate distributions in the low-velocity part and substantially different from the Maxwellian velocity and velocity rate distributions in the high-velocity part [6]. They both, as velocity goes to c, fall off to zero slower than any exponential decay but faster than any power law decay [9].

Evidently, the nuclear fusion reaction rate based on the relativistic equilibrium velocity distribution, R, has a reduction factor with respect to that based on the Maxwellian velocity distribution, $R_M$ [10],

$$R = \frac{\tanh Q}{Q} R_M, \tag{12a}$$

$$Q = (2\pi z_1 z_2 \frac{K_B T}{\mu c^2} \frac{e^2}{\hbar c})^{1/3}. \tag{12b}$$

$R_M$ is actually the first-order approximation of R. Since $0 < Q < \infty$, the reduction factor satisfies $0 < \tanh Q/Q < 1$. That gives $0 < R < R_M$. The reduction factor depends on the temperature, reduced mass, and atomic numbers of the studied nuclear fusion reactions. The relativistic equilibrium velocity distribution will lower solar neutrino fluxes and change solar neutrino energy spectra. On the other hand, since most solar ions crowd in the low-velocity part at solar temperatures and densities and this part is involved in the calculations of solar sound speeds, since in the low-velocity part no significant difference exists between the relativistic equilibrium velocity distribution and the Maxwellian velocity distribution, there will be no significantly different results of solar sound speeds when we use various distributions. In this way, standard solar models will be improved.

ACKNOWLEDGMENT

The author greatly appreciates the teachings of Prof. Wo-Te Shen. The author thanks Prof. Gerhard Muller and Dr. P. Rucker for their supports of this work.

REFERENCES

[1] J. N. Bahcall and M. H. Pinsonneault, Rev. Mod. Phys., 67, 781 (1995)

F. Ciacio, S. Degl'Innocenti and B. Ricci, Astron. Astrophys. (Suppl.), 123, 449 (1997)

A. Dar and G. Shaviv, Astrophys. J., 468, 933 (1996)

J. A. Guzik and A. N. Cox, Astrophys. J., 411, 394 (1993)

C. R. Proffitt, Astrophys. J., 425, 849 (1994)

O. Richard et al, Astron. Astrophys., 312, 1000 (1996)

[2] S. Turck-Chieze et al, Phys. Rep., 230, 57 (1993)

[3] J. N. Bahcall et al, Phys. Rev. Lett., 78, 171 (1997)

J. N. Bahcall, hep-ph/0009044

S. Basu et al, Astrophys. J., 460, 1064 (1996)

Solar Neutrino: The First Thirty Years, eds. J. N. Bahcall et al, Addison Wesley, New York (1995)

[4]   J. N. Bahcall, Neutrino Astrophysics, Cambridge University Press, Cambridge (1989)

[5]   Jian-Miin Liu, Chaos, Solitons&Fractals, 12, 399 (2001) [physics/0108044]; 12, 1111 (2001) [the revised version of hep-th/9805004]

[6]   Jian-Miin Liu, Chaos Solitons&Fractals, 12, 2149 (2001) [physics/0108045]; cond-mat/0108356; astro-ph/0108304

[7]   A. Sommerfeld, Phys. Z., 10, 826 (1909)

V. Varicak, Phys. Z., 11, 93; 287; 586 (1910)

W. Pauli, Theory of Relativity, Pergamon Press Ltd., New York (1958), trans. G. Field

V. Fock, The Theory of Space Time and Gravitation, Pergamon Press, New York (1959)

R. H. Boyer, Am. J. Phys., 33, 910 (1965)

D. K. Sen, J. Math. Phys., 31, 1145 (1990)

[8]   Jian-Miin Liu, Velocity space in the modified special relativity theory, to be published

[9]   Jian-Miin Liu, cond-mat/0112084

[10]  Jian-Miin Liu, Relativistic corrections to the Maxwellian velocity distribution and nuclear fusion reaction rate, to be published